\documentclass[12pt,onecolumn,a4paper]{article}
\usepackage{color, colortbl}
\usepackage{amsmath, amssymb}
\usepackage{graphicx}
\usepackage[T2A,T1]{fontenc}
\usepackage[utf8]{inputenc}
\usepackage[russian,english]{babel}

\definecolor{Gray}{gray}{0.9}

\begin{document}

\begin{center}
\Large{\bf Can rodents conceive hyperbolic spaces?}

\bigskip

\normalsize {\sl Eugenio Urdapilleta$^{1}$, Francesca
Troiani$^{1}$, Federico Stella$^{1}$, Alessandro Treves$^{1,2}$}

\bigskip

1: Cognitive Neuroscience, SISSA, via Bonomea 265, 34136 Trieste,
Italy\\ 2: ale@sissa.it (corresponding author).

\end{center}

\noindent {\bf Summary}

\bigskip

The grid cells discovered in the rodent medial entorhinal cortex
have been proposed to provide a metric for Euclidean space,
possibly even hardwired in the embryo. Yet one class of models
describing the formation of grid unit selectivity is entirely
based on developmental self-organization, and as such it predicts
that the metric it expresses should reflect the environment to
which the animal has adapted. We show that, according to
self-organizing models, if raised in a non-Euclidean hyperbolic
cage rats should be able to form hyperbolic grids. For a given
range of grid spacing relative to the radius of negative curvature
of the hyperbolic surface, such grids are predicted to appear as
multi-peaked firing maps, in which each peak has seven neighbours
instead of the Euclidean six, a prediction that can be tested in
experiments. We thus demonstrate that a useful universal neuronal
metric, in the sense of a multi-scale ruler and compass that
remain unaltered when changing environments, can be extended to
other than the standard Euclidean plane.

\bigskip
\bigskip

\noindent {\bf Keywords}

\bigskip

\noindent Grid cells - Self-organizing process - Space
representation - Hyperbolic geometry

\bigskip
\bigskip

\noindent {\bf Abbreviations}

\bigskip

\noindent mEC, medial entorhinal cortex; PS, pseudosphere

\newpage
\section{Introduction}
\indent Euclidean geometry was long suspected not to be the sole
possible description of physical space (e.g. by Omar Khayyam in
his 1077 book \textit{Explanations of the difficulties in the
postulates in Euclid's Elements} \cite{Khayyam1077,
Rozenfeld1988}). It was only around 1830, however, that rigorous
non-Euclidean alternatives were formulated independently by
Nikolai Lobachevski (first communicated on February 23, 1826, then
printed in a Russian journal
%{\fontencoding{T2A} \selectfont
%\char194 \char229 \char241 \char242 \char237 \char232 \char234}
%{\fontencoding{T2A} \selectfont \char202 \char224 \char231
%\char224 \char237 \char241 \char234 \char238 \char227 \char238}
%{\fontencoding{T2A} \selectfont \char243 \char237 \char232
%\char226 \char229 \char240 \char241 \char232 \char242 \char229
%\char242 \char224}
in 1829) and by J\'anos Bolyai (between
1820-23, but only published in the Appendix to a 1831 textbook by
his father) \cite{Bonola1912}. The great Friedrich Gauss wrote
that he had been thinking along similar lines, but recognized the
genius of his younger colleague (e.g. in his letter to Gerling, in
1832) \cite{Halsted1900}. Given the involvement of such profound
thinkers and the centuries intervened, it makes sense to ask
whether the development of non-Euclidean formulations was
intrinsically arduous, even for brilliant individuals, or rather
it was made difficult by stratified scientific conventions and
conformist patterns of thought - a social group effect.

An unusual approach to this question is provided by the discovery,
in the rodent brain, of grid cells \cite{Fyhn_etal2004,
Hafting_etal2005}. These neurons, with their activity concentrated
at the nodes of a surprisingly regular triangular grid, which is
different from neuron to neuron, appear similar to sheets of graph
paper, lining the environment in which the animal moves
\cite{Moser_etal2008}. Indeed such cells have been proposed to
comprise a Euclidean metric of physical space \cite{Solstad2009},
providing a single common population gauge to measure the
environment \cite{Fyhn_etal2007, MoserMoser2008}. But is it
necessarily Euclidean? If non-Euclidean grid cells were discovered
in rodents, still sufficiently regular to be characterized as
providing a metric of space, it would indicate that the human
difficulty at conceiving non-Euclidean geometry is likely a social
group effect.

Grid cell activity has been mostly studied in experiments in which
rodents forage in flat, horizontal, open 2D environments
\cite{Hafting_etal2005, Barry_etal2007, Stensola_etal2012}. Grid
units have been also observed (also in flat arenas) in mice
\cite{Fyhn_etal2008} and crawling bats \cite{Yartsev_etal2011}.
Some experiments have utilized apparatuses that probe the vertical
dimension \cite{Hayman_etal2011}, but the resulting observations
do not seem to point at a simple abstraction of general validity
(see discussion in \cite{Jeffery_etal2013}). The other major type
of spatially selective units, the place cells, has been observed
also in flying bats, and they appear to span 3D space fairly
isotropically \cite{YartsevUlanovsky2013}. Grid cells may yet be
observed in flying bats - there have been preliminary observations
of ``blobby'' i.e. multi-peaked cells \cite{Ulanovsky_Fens2014},
and if their periodicity is confirmed this would invoke models of
grid formation that can function in 3D, not just 2D
\cite{Stella_etal2013a}. In order to address the uniqueness and
necessity of an Euclidean metric representation, however, one may
focus on 2D environments, and consider surfaces that can be
described by a regular, position-invariant non-Euclidean metric.
From the activity observed in grid units, one should be able to
infer whether the rat brain can only express an orderly
representation of a Euclidean space, or it can also adapt to an
environment defined by a non-Euclidean metric.

Non-Euclidean position-invariant 2D metrics can be exemplified by
surfaces embedded in physical 3D space, of two types: those with
constant positive Gaussian curvature, i.e. spheres or portions of
spheres, and those with constant negative curvature, the so-called
pseudospheres introduced by Beltrami \cite{Beltrami1868} (see
figure 1). The former is a model of elliptic geometry, the latter
of hyperbolic geometry. In \cite{Stella_etal2013b} we have
analysed the prediction of a particular self-organizing model of
grid cell development \cite{KropffTreves2008}, and concluded that
rats exploring a spherical cage of appropriate radius should
develop grids with coordination number 5 or lower, i.e. each of
their peaks should be surrounded by 5 (or fewer) nearest
neighbours instead of the Euclidean 6. Experiments are underway to
examine grid cells in rats raised in a cage that closely
approximates a sphere \cite{Kruge_etal2013}.

Still, spherical geometry departs from Euclidean geometry in other
major ways than just being non-flat: it has no points at infinity
(and so no indefinitely long geodesic trajectories, whether
parallel or not, nor much of a requirement for path-integration
along such trajectories); and a complete sphere has no boundary,
eliminating the need and the opportunity of discontinuities in the
representation at the boundary (although one can cut an artificial
boundary, or use a portion of a sphere). So, should the grid
cells, which emerge in rats raised in or adapted to a sphere,
reveal the striking soccer-ball pattern predicted by our model,
the effect of a non-Euclidean metric might be argued to be
confounded with that of the missing points at infinity, for
example. Therefore we aim here to consider the emergence of grid
units in rats raised in a hyperbolic cage, as a more stringent
potential test of the ability of rodents to conceive non-Euclidean
representations of space.

A similar asymmetry between elliptic and hyperbolic geometry was
perceived by Hieronymus Saccheri, who  in his Euclides Vindicatus
(1733) tried to prove Euclidean geometry by refuting its two
alternatives \cite{Saccheri1733}; but while he had an easy task
dismissing elliptic geometry, because it would violate also
Euclid's second postulate by not having infinitely long straight
lines, he had difficulties refuting hyperbolic geometry. He wrote
of the ``difference between the foregoing refutations of the two
hypotheses. For in regard to the hypothesis of the obtuse angle
[elliptic] the thing is clearer than midday light ... But on the
contrary, I do not attain to proving the falsity of the other
hypothesis, that of the acute angle [hyperbolic] ... I do not
appear to demonstrate from the viscera of the very hypothesis, as
must be done for a perfect refutation''.  It makes sense to ask
whether the brain of a rat exploring a hyperbolic environment can
demonstrate what the ``viscera of the very hypothesis'' failed to
refute.

\section{Results}
\subsection{Pseudospheres can be used to test the emergence of
non-Euclidean neural representations}

The idea of raising rats in a cage with a hyperbolic shape, and
checking what kind of grid units, if any, they develop in their
medial entorhinal cortex (mEC), runs into two difficulties, if the
cage is realized as a simple (half) pseudosphere (PS), see figure
1{\sl a}. The first difficulty stems from the circular symmetry of
the PS around its z-axis, and the second from its limited area.

\begin{figure*}[!ht]
\begin{center}
\includegraphics[scale = 1, angle = 0]{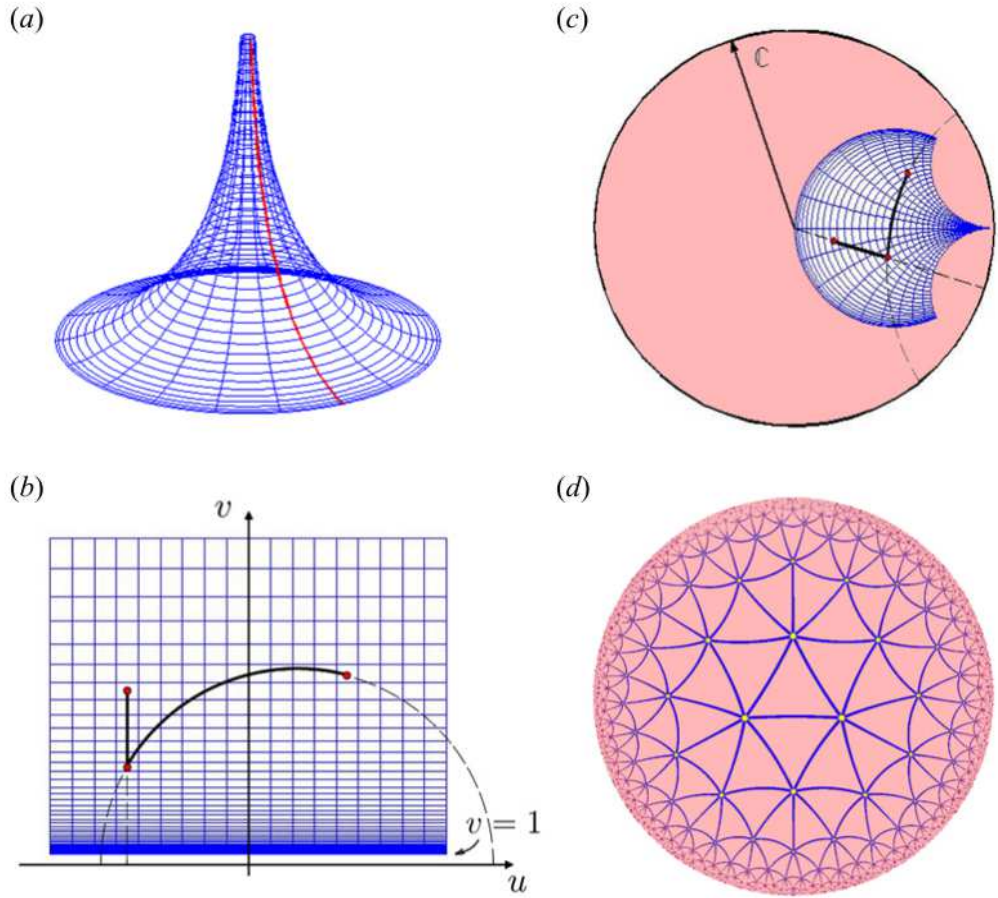}
\caption{A 2D model of hyperbolic geometry. ({\sl a}) The (half)
PS is generated by revolution of the tractrix (red curve) around
the $z$-axis. ({\sl b}) Representation of the PS in the Poincar\'e
half-plane. ({\sl c}) A M\"obius transformation maps it into the
Poincar\'e disk. ({\sl d}) Its constant curvature allows for
regular tessellations, starting from heptagonal ones, with a dual
grid visualized here in the Poincar\'e disk. (Online version in
colour.)}\label{fig1}
\end{center}
\end{figure*}

The circular symmetry problem can be understood by mapping the PS
into the Poincar\'e half-plane (figure 1{\sl b}) and into the
Poincar\'e disk (figure 1{\sl c}). These are two flat
representations of an infinitely extended 2D hyperbolic space, in
which a constant negative Gaussian curvature is expressed instead
by assigning a local metric which is a function of position (see
Methods). The revolution $-\pi \leq \theta \leq \pi$ generating
the PS makes its geodesics $\theta=-\pi$ and $\theta=\pi$
coincide, but the corresponding geodesics do not coincide on the
half-plane or on the disk. As a result, any regular infinite
tiling (see, for example, figure 1{\sl d}), which has no reason to
be periodic e.g. on the $u$ variable of the half-plane, once cut,
transformed and pasted onto the PS will be discontinuous at
$\theta=\pm\pi$. Viceversa, any continuous grid map developing in
a rat freely roaming around the PS will be incompatible with a
regular tiling. The simple solution to this difficulty is to
insert a partition at $\theta=\pm\pi$, making it impossible to
walk around the PS. This solution can be applied both to an
experimental setting and to computer simulations.

The half PS has a finite area $2\pi R^2$ and it can be mapped only
onto a limited portion of the half-plane or of the disk (figures
1{\sl b}, 1{\sl c}). Since its area is limited, it cannot support
many tiles. For a regular triangular tiling with $q$ triangular
tiles meeting at each node, i.e. with angles $2\pi/q$, each tile
has area $\pi(1-6/q)R^2$ i.e. a fraction $(1-6/q)/2$ of the total
available (see Methods). The dual polygonal tile centered at each
node includes $1/3$ of the area of $q$ triangles, hence takes up
an area $N_q = q \pi (1-6/q)R^2 / 3$. Therefore the half PS can
include on its finite area the equivalent of $2 \pi R^2 / N_q =
6/[q(1-6/q)]$ nodes, which gives $\infty$ for $q=6$ (i.e. the
hexagonal dual tiles would be infinitely small relative to the
curvature, in practice populating a Euclidean plane); $6$ for
$q=7$ (heptagonal grids); $3$ for $q=8$ (octagonal grids); $2$ for
$q=9$; $1$ for $q=12$; etc. Hence beyond $q=7$ the possibility to
see a symmetric grid tiling on a half PS vanishes rapidly, because
too few are the fields that each unit could fit on the finite
surface. The spacing of a symmetric grid can be calculated to be
\begin{equation}
  l = R \, \text{cosh}^{-1}\left[ \frac{\gamma}{1-\gamma}\right],
\end{equation}

\noindent where $\gamma=\cos(2\pi/q)$, which gives infinitesimal
$l$ for $q=6$, as it should; $l\approx 1.09\,R$ for $q=7$;
$l\approx 1.53\,R$ for $q=8$; $l\approx 1.85\,R$ for $q=9$; etc.

\begin{figure*}[!tp]
\begin{center}
\includegraphics[scale = 1, angle = 0]{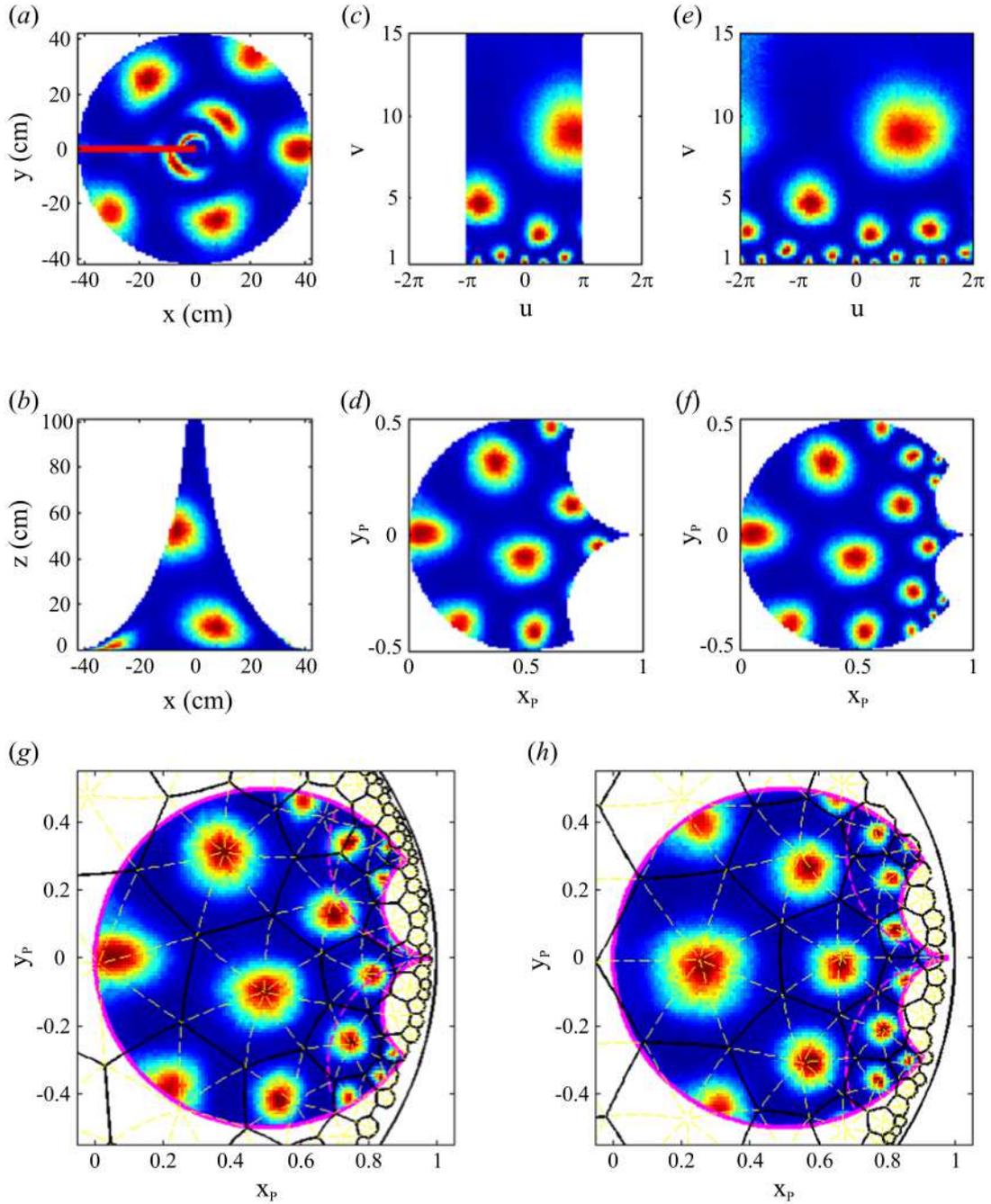}
\caption{An approximately heptagonal grid self-organized in
simulations. The PS is seen ({\sl a}) from above and ({\sl b})
from the side, projected ({\sl c}) on the half-plane and ({\sl d})
on the Poincar\'e disk, and shown with the added folds in ({\sl
e}) and ({\sl f}). ({\sl g}) An exact heptagonal tessellation
describes well the multi-peaked activity of the example unit shown
in ({\sl a}-{\sl f}). ({\sl h}) The heptagonal grid of another
unit. (Online version in colour.)}\label{fig2}
\end{center}
\end{figure*}

In simulations, one can deal with the limited area of the PS by
adding lapels or folds on each side of the corresponding portion
of the Poincar\'e half-plane. We limit ourselves to doubling the
area, thus making space for $12$ fields for perfectly heptagonal
grids, $6$ for octagonal, and so on. With parameters that yield
the appropriate grid spacing, approximately heptagonal grids are
easily self-organized (figure 2).

\subsection{The symmetry of the emerging grid units reflects
Gaussian curvature}

In the self-organizing model we use, the mean grid spacing is
determined by the adaptation time scale, which once multiplied by
the average exploration speed becomes a length scale. The relation
between the mean grid spacing and the radius of curvature selects
the type of tiling that emerges. An exact heptagonal or octagonal
tiling would require the spacing-to-radius ratios reported above.
Interestingly, when the ratio takes intermediate values, one
observes in simulations the emergence of grids with intermediate
mean angle between triplets of spikes in neighbouring fields
(figure 3; and see Methods for the procedure used to measure such
angles).

\begin{figure*}[!ht]
\begin{center}
\includegraphics[scale = 1, angle = 0]{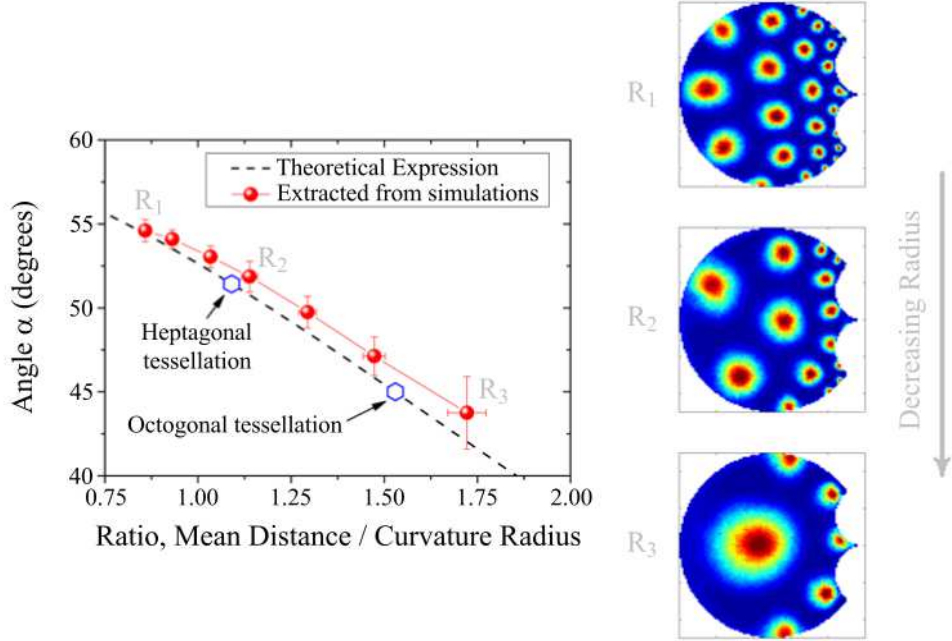}
\caption{Left: A portion of the universal curve relating the
angles of equilateral triangles to their linear size in relation
to the radius of curvature of a regular surface (dashed line). The
average angle between triplets of spikes belonging to neighbouring
fields, in grid units emerging from simulations, falls close to
the universal curve (red symbols). Exact tessellations correspond
to those regular triangles that tile space indefinitely (blue
empty symbols). Right: Decreasing the radius of the folded PS,
equivalent to an increase in curvature, changes the angular
relationships between neighbouring fields. (Online version in
colour.)}\label{fig3}
\end{center}
\end{figure*}

\subsection{Population coherence induced by lateral interactions}

So far, each grid unit has developed independently. Shared single
unit properties define a common grid spacing, but grid orientation
is randomly distributed. To induce a common orientation, we now
add a collateral network where the connection strength between any
two units is determined by the virtual position procedure
described by \cite{Si_etal2012}. This interacting network produces
conjunctive head direction x grid units, mimicking the cells found
in intermediate and deep layers of mEC \cite{Sargolini_etal2006,
SiTreves2013}.

To assess the effect of the collaterals on our hyperbolic grids,
we analyze the firing activity of each unit in the half-plane, see
figure 4{\sl a}. Since there translations and rotations are
coupled, we cannot simply shift and collapse the activity of all
units in a common origin, and evaluate the resulting common
alignment. Instead, we focus on the activity of multiple single
units in different portions of the plane, eg in the yellow box in
figure 4{\sl a}. For each spatial bin on the half-plane, we
average across the population the weighted distribution of angles
between pairs of spikes, one belonging to that bin and the other
to a surrounding annular region (see Methods). The weight is
determined by the number of spikes of each unit in the bin itself,
so clear peaks emerge in the angular distribution if all units
having a field in the bin share the location of neighbouring
fields. For example, in the bin indicated by the yellow box, the
unit shown in figure 4{\sl a} has the distribution of angles
represented in figure 4{\sl b} as a black line. The population
distribution, represented by the gray line in figure 4{\sl b},
preserves the multi-peaked structure of individual units,
indicating population coherence. In contrast, the distribution of
angles for pseudo-spikes randomly allocated to visited locations
(red dashed line) is less peaked, as it is the one for spikes
shuffled across the population (blue dashed line). The agreement
between these two distributions, which holds for any bin,
indicates in fact that different units sample the PS evenly,
expressing a distributed representation. To quantitatively assess
the influence of the collaterals we take the integral of the
square difference between the population distribution of angles
and either of these two control distributions. By summing this
square difference over all bins of equal hyperbolic area on the PS
(except those at the boundary), we get a summary coherence
measure. Whatever the PS curvature, we observe an increase in the
coherence measures when adding the collaterals (moderate with the
random control, fourfold with the shuffled control), showing that
they contribute significantly to align the population response
(Table 1).

\begin{figure*}[!ht]
\begin{center}
\includegraphics[scale = 0.95, angle = 0]{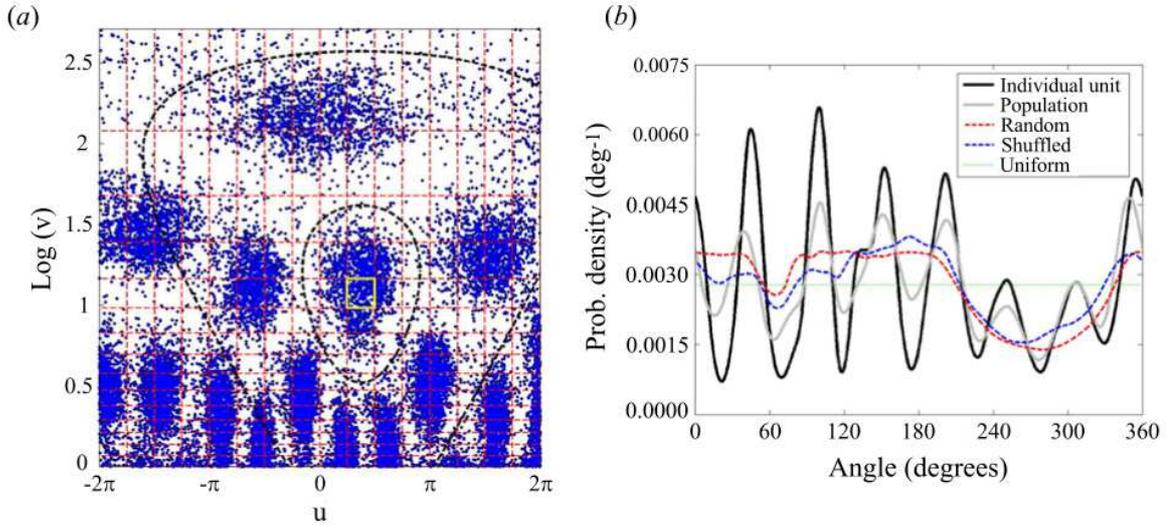}
\caption{Hyperbolic grid alignment by collaterals. ({\sl a}) The
PS in the half-plane is divided in a regular partition. For a
given spatial bin, eg the yellow box, spikes from individual units
are considered, paired with those located at a certain distance
matching grid spacing. ({\sl b}) The distribution of angles
between these spikes, measured relative to the spike in the chosen
bin, are calculated for each unit (black line), and accumulated in
a weighted average over the population (gray line). Similar
distributions are obtained from randomly positioned pseudo-spikes
(random, red dashed line) or shuffled across the population
(shuffled, blue dashed line). The integral square differences
between the population distribution and these controls are used to
quantify the effect of the collaterals on grid alignment (Table
1). (Online version in colour.)}\label{fig4}
\end{center}
\end{figure*}

\begin{table}[!ht]
\begin{center}
\begin{tabular}
{lcc} PS radius & Random control & Shuffled control
\cr\hline\hline $R = 35$ cm & $0.0171\, \text{deg}^{-1}$ &
$0.0060\, \text{deg}^{-1}$\cr \rowcolor{Gray} $R = 35$ cm &
$0.0294\, \text{deg}^{-1}$ & $0.0259\, \text{deg}^{-1}$\cr \hline
$R = 40$ cm & $0.0192\, \text{deg}^{-1}$ & $0.0051\,
\text{deg}^{-1}$\cr \rowcolor{Gray} $R = 40$ cm & $0.0263\,
\text{deg}^{-1}$ & $0.0189\, \text{deg}^{-1}$\cr \hline $R = 45$
cm & $0.0160\, \text{deg}^{-1}$ & $0.0039\, \text{deg}^{-1}$\cr
\rowcolor{Gray} $R = 45$ cm & $0.0272\, \text{deg}^{-1}$ &
$0.0189\, \text{deg}^{-1}$\cr \hline $R = 50$ cm & $0.0166\,
\text{deg}^{-1}$ & $0.0038\, \text{deg}^{-1}$\cr \rowcolor{Gray}
$R = 50$ cm & $0.0196\, \text{deg}^{-1}$ & $0.0145\,
\text{deg}^{-1}$\cr \hline
\end{tabular}
\caption{Coherence measure in a population
with/without collaterals. Single unit properties define a grid
spacing of approximately $45$ cm, slightly modified when the
collaterals are added. Rows in gray correspond to the system with
collaterals.}
\end{center}
\end{table}

\subsection{Planar grids can re-adapt coherently to hyperbolic spaces}

To bridge the gap between model predictions and experimental
conditions suitable to observe heptagonal tessellations, we first
consider what happens if rats are raised in planar environments
and then experience a hyperbolic one. Given single-unit properties
that yield the proper grid spacing on the PS and assuming
continued plasticity of the feed-forward connections, simulations
indicate that individual units rapidly develop a local structure
compatible with a heptagonal grid (not shown). The long-range
structure as well as the common alignment of the population
require more adaptation, but the timescale is not longer than when
the model adapts to a second planar environment
\cite{Si_etal2012}. Interestingly, as with the adaptation to a
second planar environment (which reduces to a translation and
rotation of the stack of grid maps, cp. \cite{Fyhn_etal2007}),
also in this more complex case the original phase relations
between any two units are kept in the second environment, though
this is now a hyperbolic PS. Therefore the collaterals, while weak
enough as to not interfere with heptagonal grid development, can
be sufficient to maintain or re-learn the phase relationships in
the new environment.

To show this, we consider the cross-correlation of the activity
developed in the planar environment between all pairs of units in
the population, see figures 5{\sl a1}-{\sl a3}. The peak closest
to the center in each cross-correlogram defines the phase between
the two units and, as shown in figure 5{\sl a}, it is broadly
distributed but with a tendency to form clusters. Based on this
distribution, we can classify each pair according to the region
where their relative phase is located, see the white dashed lines
delimiting regions {\sl a1}-{\sl a3} in figure 5{\sl a}. The
correlation for each of these pairs can be then measured after
adaptation to the PS, based on their activity in the half plane as
in the previous analysis, and contrasted with their planar
correlation. As observed in figure 5{\sl b}, the two correlation
values are related, and respect the classification in figure 5{\sl
a} (different sets {\sl a1}-{\sl a3} are indicated by colors).
Representative examples of out-of-phase and in-phase relationships
are shown in figure 5{\sl b1} and figure 5{\sl b2}, respectively.

\begin{figure*}[!ht]
\begin{center}
\includegraphics[scale = 0.95, angle = 0]{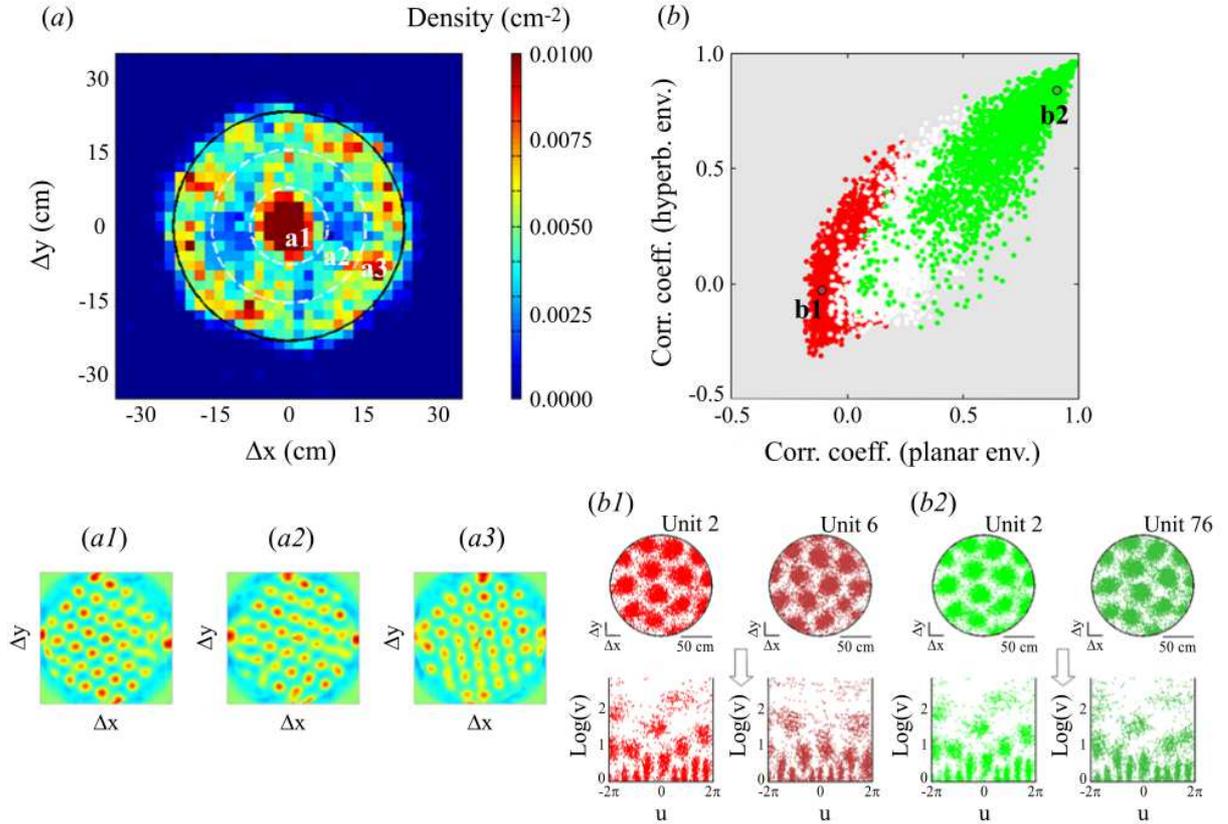}
\caption{Planar phase relationships are maintained on the PS.
({\sl a}) Distribution of the peaks closest to the origin in the
cross-correlograms obtained from all pairs of units in an initial
planar environment. Three phase relation categories are defined,
denoted {\sl a1}-{\sl a3}. Representative examples of their
cross-correlations are shown below. ({\sl b}) Pearson correlation
coefficient between pairs of units obtained from the activity they
develop on a PS, as a function of their correlation coefficient in
the original planar environment. Colors indicate the
classification in ({\sl a}): Green/White/Red - {\sl a1}/{\sl
a2}/{\sl a3}, respectively. Examples of ({\sl b1}) out-of phase
and ({\sl b2}) in-phase relationships are shown below, in both
environments. (Online version in colour.)}\label{fig5}
\end{center}
\end{figure*}

\subsection{A rodent-friendly hyperbolic box}

The previous analyses indicate the possibility of observing
hyperbolic properties in the activity of grid cells, if the
self-organization model is valid. However, to double its area and
so reliably detect a heptagonal tessellation, we resorted to
folding the PS around itself. Such a procedure makes no
experimental sense. A second approach is to increase the available
area by a similar amount adding a surface that, \textit{on
average}, has a similar hyperbolic curvature, but not constant in
its negative value. This can be realized by a sort of wavy
continuation, see figure 6{\sl a}. As seen in figures 6{\sl
b}-6{\sl e}, a network of units with suitable properties again
develops, through exploration of this extended surface, a grid
representation with the characteristic coordination number $7$
and, at least over the central pseudospherical portion, equivalent
regularity as for the folded PS.

\begin{figure*}[!ht]
\begin{center}
\includegraphics[scale = 1, angle = 0]{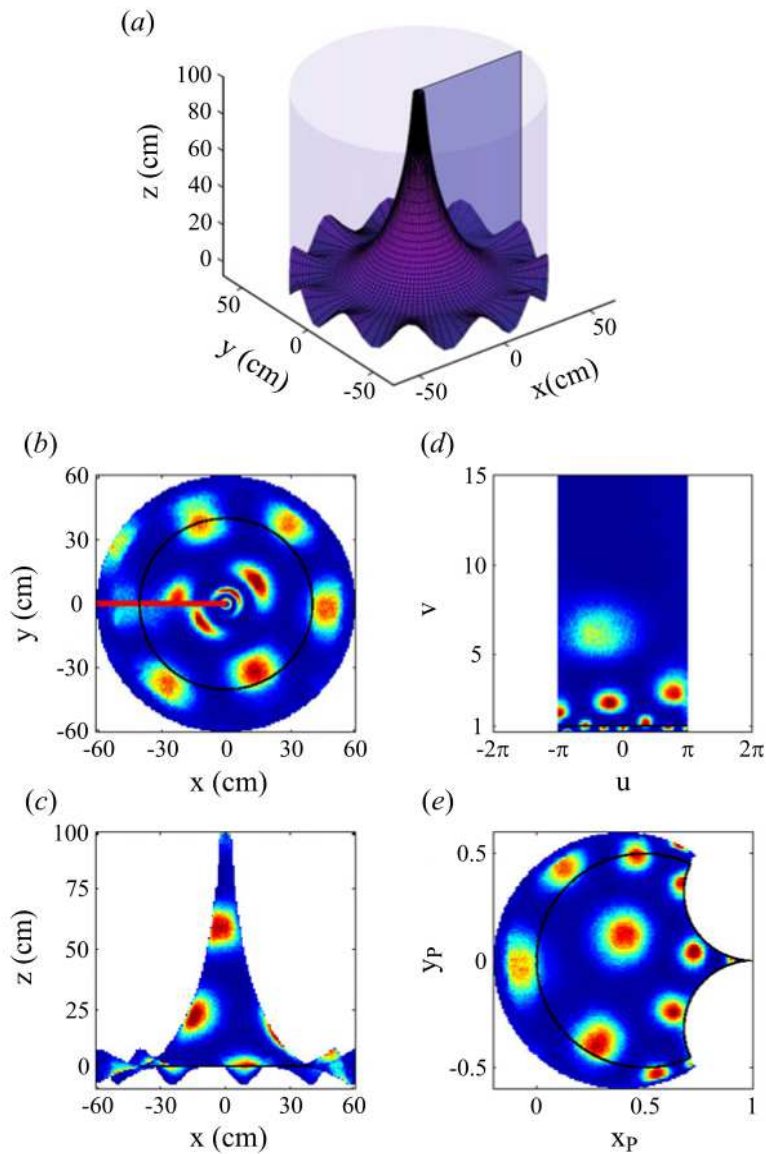}
\caption{A plausible environment for experiments with rodents.
({\sl a}) The non-folded PS with a wavy continuation. Activity
developed in a grid unit, seen ({\sl b}) from above and ({\sl c})
from the side. Projection onto ({\sl d}) the half-plane and ({\sl
e}) the Poincar\'e disk. Black lines indicate the continuation
beyond the standard PS. (Online version in colour.)}\label{fig6}
\end{center}
\end{figure*}

\section{Discussion}

We have shown that in a self-organizing model, grid units adapt
effortlessly to hyperbolic geometry, and given appropriate values
of Gaussian curvature can express maps with coordination number 7.
If the model captures the essential character of grid formation,
one expects to be able to observe heptagonal grids also in rodents
adapted to quasi-regular extensions of the basic pseudospherical
surface. The model further indicates that while hyperbolic
adaptation has to be protracted, it can be subsequent to an
earlier phase of grid development in an ordinary planar
environment. Critical to this outcome is the prolonged plasticity
of the relevant synaptic connections, which in the model are those
afferent to the developing grid units.

It is likely, but it remains to be established conclusively, that
the transition or ``remapping'' from planar to hyperbolic maps is
facilitated, with respect to the planar-to-spherical case, by
topological similarity: the planar arena and the PS, whether
folded or extended, are limited compact portions of an infinite
space, $R^2$ or $H^2$, surrounded by a boundary (provided the PS
includes the partition); the sphere has no boundary if complete
and it can include, even if not complete, closed geodesics along
which an activity map has to be matched with itself.

Do alternative models of grid cell formation predict heptagonal
grids? It appears not to be the case for oscillatory interference
models, at least in their original formulations
\cite{Burgess_etal2007, Blair_etal2007}, which are based on the
superposition of three cosine waves. For attractor network models,
our own simulations indicate that the same collateral network
which aligns and sustains planar grids can maintain their phase
relationship also on a PS. Such models, however, do not fully
specify the process which leads to the emergence of the grid
pattern in the first place, other than invoking an embryonal
position of the neurons in the tissue which determines the
collateral connectivity and the phase relations, and is then
somehow transformed to become unrelated to the phase relations in
the adult animal \cite{McNaughton_etal2006}. The connectivity set
up in the embryo would have to be slightly hyperbolic, but it
could well be, and in any case at the relevant scale of the unit
tile of the tessellation the difference between planar and
hyperbolic might be negligible. Therefore attractor models are
compatible with hyperbolic grid formation, but do not really
predict it, in the same sense that they are under-formulated to
predict planar grid formation.

Several open questions about the self-organizing models apply to
their development in hyperbolic geometry as well. These include
the delayed-action mechanism that leads to the self-organization
of the recurrent connections, the role of the layered structure
\cite{SiTreves2013}, the interactions with hippocampal and lateral
entorhinal cortex cells. On the other hand, the regularity of the
activity pattern expressed by these self-organized grid units
relies on the assumption of an even coverage of the available
surface. Whereas behaviorally this can be easily achieved on flat
surfaces, eg by stimulating the rat to chase chocolate chips
around the arena, the even exploration of curved surfaces under
gravity can be experimentally challenging. These issues deserve
further analysis.

Would the discovery of hyperbolic representations in rodents bear
implications for human cognition, beyond the suggestion that our
fellow mammals may be less inhibited by social conventions? We
reckon that such a finding would have to be taken into account in
the fascinating analysis of what are often called {\em spatial
primitives}, either in infants \cite{Dillon_etal2013} or in
indigenous tribes who may or may not have been able to elude, so
far, geometrical globalism \cite{Dehaene_etal2006,
Izard_etal2011}.

\section{Methods}
\subsection{Models of hyperbolic surfaces}
It is not possible to represent, in physical 3D space, the full
hyperbolic $H^2$ space of constant negative Gaussian curvature
$-1/R^2$. At any point on such a mathematically defined surface
one can identify two orthogonal axes, along which the surface
would appear to curve on opposite sides of the locally tangent
Euclidean plane, with $R^2$ as the product of the two radii of
curvature. Then one can resort to models, giving up one or another
of the features of a true representation of the full $H^2$.

The PS or tractricoid is a non-distorted, 3D representation of a
finite portion of $H^2$, which is useful in that both distances
(measured along geodesics, the minimal length paths between any
two points on the surface) and angles are preserved. As a result,
no speed transformation is needed for an animal who moves on a PS
to experience hyperbolic metricity, and this makes the PS in
principle suitable as the design of a hyperbolic rat cage. The
main limitation is that while the PS has two points at infinite
distance (the cusps) it has only finite area $4\pi R^2$. In fact,
it is comprised of two halves joined at a base (a circumference of
radius $R$ in 3D space) which cannot be crossed with continuous
movements, so for all practical purposes we always consider here a
half PS, with area $2\pi R^2$ and a single cusp, that we take to
point in the vertical direction $z$ of physical 3D space. The
position of any point on the half PS can be given by its 3D
distance $r$ from the $z$-axis, and by its angle $\theta$ from an
arbitrarily chosen reference direction. The height $z$ of the
point is related to $r$ as $z(r)= R\,[\text{cosh}^{-1}(R/r) -
\sqrt{1-(r/R)^2}]$ and the infinitesimal curve element has length
$ds$, best expressed in terms of the variable $v=R/r$ as
$ds^2=[d\theta^2+dv^2]/v^2$. One can view the half PS from above,
projected on a circle of radius $R$ on the plane (since $r<R$),
but at the price of distorting the local metric in the projection.
It is useful then to consider other 2D models which, while also
distorting the local metric and thus impairing appreciation of
distances and angles, at least can represent an infinitely
extended $H^2$.

The Poincar\'e half-plane model amounts simply to unwrapping the
PS around the $z$-axis, stretching its surface near the cusp and
plotting $\theta$ (often denoted as $u$) and $v$ as Cartesian
variables, which however can now extend $-\infty < \theta <
\infty$ and $0 < v < \infty$. The local metric is defined as
$ds^2=(du^2+dv^2)/v^2$ and its geodesics are either vertical lines
$u=$const or circular arcs centered on the $v=0$ line (see figure
1{\sl b}). The PS corresponds to the ranges $-\pi < \theta < \pi$
and $1 < v < \infty$ on the half-plane. One has to remember that
the height $z$ of a point on the PS grows with its corresponding
height in the half-plane, but via the non-linear transform $z(v)=
R\,[\text{cosh}^{-1}(v) - \sqrt{1-(1/v)^2}]$.

The Poincar\'e disk is another distorted two-dimensional
representation of half the hyperbolic plane of constant negative
curvature $H^2$, which is useful in particular to display its
possible regular triangulations, in which a node is neighbour to
$q = 7, 8,\dots, \infty$ other nodes (in Euclidean space the only
regular triangular tiling has $q = 6$, as for grid cells, while on
a spherical surface $q = 5, 4, 3, 2$ or $1$
\cite{Stella_etal2013b}). The disk of coordinates $(x_P,y_P)$,
with $x_P^2+y_P^2<1$, has a local metric defined as $ds^2 =
(dx_P^2+dy_P^2) / [1-x_P^2-y_P^2]^2$ and its geodesics are
diameters of the disk, or circular arcs that meet its boundary
orthogonally (see figure 1{\sl c}).

The PS, given by the revolution of the so-called tractrix around
the $z$-axis, can be also defined as the surface spanned by the
cylindrical coordinates $(r,\theta,z)$ where, with reference to
the Poincar\'e half-plane, $\theta=u$ while $r$ and $z$ are
parameterized as $r = R\,\text{sech}(t)$ and
$z=R\,[t-\text{tanh}(t)]$, with $t=\text{cosh}^{-1}(v)$ and $v\geq
1$. The half-plane and the disk are isomorphic and can be related
by a M\"obius transformation, for example by $x_P = (u^2+v^2-1) /
[u^2+(v+1)^2]$ and $y_P = -2u / [u^2+(v+1)^2]$, which maps the
origin $(0,0)$ of the half-plane on the boundary $(-1,0)$ of the
disk, $(0,1)$ onto the origin $(0,0)$ of the disk, and the point
at infinity $(0,\infty)$ on the boundary $(1,0)$ of the disk. We
shall use this particular transformation in the figures. Although
its appearance is also that of a circle, like a PS seen from
above, the mapping is much more complicated, almost intuitively
inverted, with the cusp of the PS mapped on one arbitrary point on
the outer circumference of the disk, and the circumference at the
base of the PS curving through the center of the disk (figure
1{\sl c}).

\subsection{Pseudosphere + skirt}
One cannot use folds with rats, but to extend the area of the PS
one can add a ``skirt'' around it, so that additional fields can
form on the skirt, see figure 6{\sl a}. Since the skirt cannot
have a regular hyperbolic metric, one has to consider the effect
of attaching it to a surface of constant negative curvature. A
first option could be a flat skirt, eg extending up to a radius
$R'=3/2R$. The disadvantage is that fields on the skirt may tend
to flatten also the geometry on the inner PS. A second option is
to add an undulated skirt, with cosine waves extending in the $z$
direction by an amount $f(r)$ which depends on $r$ and tends to
$0$ for $r\rightarrow R$, so that $(x,y,z) = [r\,\sin
\theta,r\,\cos \theta,f(r)\,\cos(m\theta)]$. By choosing $m = 11$,
we try to minimize the bias between coordination number 6 or 7 or
8, as the 11 waves on the skirt are incommensurable with all these
patterns.

To construct a skirt that smoothly extends the PS with $m = 11$
waves, as in figure 6{\sl a}, one can compute the curvature along
each of the ``ridges'', where the vector normal to the surface, in
the direction $w$, is tilted with respect to the $z$-axis but only
along the radius. Taking the ridge to be at $y_0$ along the
$y$-$z$ plane, i.e. at $x=0$, for small deviations from
$(0,y_0,f(y_0))$ one has, according to the second fundamental
form, $w\,\sqrt{1+(f'(y_0))^2} = (x^2/2)\,[f'(y_0)/y_0 - m^2
f(y_0)/y_0^2] + [(y-y_0)^2/2]\,f''(y_0)$ so that the two
curvatures are $\kappa_y = f''(y_0) / [1+(f'(y_0))^2]^{3/2}$ and
$\kappa_x = [f'(y_0)/y_0 - m^2 f(y_0)/y_0^2] /
[1+(f'(y_0))^2]^{1/2}$, which are not constant. The ``radial''
curvature $\kappa_y$ can be positive definite and not vary much if
$f'(y_0)$ is small and $f''(y_0)$ constant, while the
``transverse'' curvature $\kappa_x$ changes sign, and the local
surface is hyperbolic only when $f'(y) \,y < m^2 f(y)$ (that is,
not at the border with the PS). A simple choice is to set
$f(r)=\sqrt{2}(r-R)/m+(r-R)^2/(2R)$. This yields along the ridge
$\kappa_y$ and $\kappa_x$ that scale like $1/R$, the first
slightly lower and the second changing sign and reaching out at
the border values close to $-10/R$. At the outer border, the skirt
is $f(3/2R)\approx 0.2R$ high.

\subsection{Model and simulation procedure}
We refer to \cite{Si_etal2012} for details, but in brief a virtual
rat is simulated to randomly explore the hyperbolic environment
described in the main text, with constant speed $40$ cm/s. At each
step, the change in the running direction is sampled from a
Gaussian distribution with zero mean and angular standard
deviation $\sigma_{\rm rdp} = 0.2$ radians - if the rat is running
on the PS - and from a Gaussian with zero mean and angular
standard deviation $\sigma_{\rm rds} = \sigma_{\rm rdp} -
0.03\,\sin \vartheta - 0.14\,(1-v)\,\sin \vartheta$ - if the rat
is running on the skirt - where $\vartheta$ and $v$ are the
direction (relative to the tangential direction, or to the
$u$-axis in the half plane) and the position of the previous step,
respectively. If the chosen direction leads the rat outside the
limits of the environment, the new position is computed by
reflecting it with respect to the crossed boundary.

The position of the virtual rat is reflected in the activity of an
input layer of place units, that feed into the output layer of
would-be grid units. The crucial self-organization occurs via
competitive learning on the feedforward connections from place to
grid units, which is modulated by recurrent connections among the
grid units. These later connections are given by an explicit rule,
as in \cite{Si_etal2012}, with a strength that grows gradually
during the initial stages of the learning process. In the
simulations reported here, the exploration and learning phases
lasted $25\times10^6$ time steps, thought to correspond to roughly
$70$ hours of real time.

In detail, the overall input to unit $i$ at time $t$ is given by
\begin{equation*}
  h_i^t = f_{\theta_i}(\vartheta^t)\,\left(\sum_j W_{ij}^{t-1} \,
  r_j^t + \rho^t \sum_k \tilde{W}_{ik} \, \Psi_k^{t-\tau}\right),
\end{equation*}

\noindent where $\Psi_k^{t-\tau}$ is the activity of unit $k$
reverberated from collateral connections $\tilde{W}_{ik}$ with a
delay $\tau = 25$ time-steps ($10$ ms each), if these connections
are present. The relative strength of the collateral inputs is set
by the factor $\rho^t$, which linearly increases from zero to a
final stationary value $0.2$ reached at half of the total
simulation time $T$, mimicking a simulated annealing. $r_j^t$ is
the firing rate of a ``place unit'' $j$ relayed by the feedforward
connection $W_{ij}$ \cite{Tonnevie_etal2013, Rowland_etal2013}.
The activity of a place unit is approximated by a Gaussian
function centered in its preferred firing location
$\vec{x}_{j_0}$,
\begin{equation*}
  r_j^t = \exp\left(-\frac{|\vec{x}^t - \vec{x}_{j_0}|^2}{2\sigma_p^2}\right),
\end{equation*}

\noindent where $\vec{x}^t$ is the current location of the
simulated rat, $\sigma_p = 5$ cm is the width of the firing field,
and $|\cdot|$ is the distance in the half plane - if the rat is on
the PS - and the distance given by the metric of the skirt - if
the rat is on the skirt -. Place cells are orderly located on the
surface, separated by approximately $5$ cm from each other. In the
network with collaterals, each unit $i$ is arbitrarily assigned
with a preferred head direction $\theta_i$ to modulate its inputs.
$f_{\theta_i}(\vartheta^t)$ is a tuning function that produces a
maximum output when the current head direction $\vartheta^t$ is
along the preferred direction $\theta_i$ \cite{Zhang1996}:
\begin{equation*}
  f_{\theta}(\vartheta) = c + (1-c)\,\exp \left\{ \nu \left[
  \cos(\theta-\vartheta)-1 \right] \right\},
\end{equation*}

\noindent where $c = 0.2$ and $\nu = 0.8$ are parameters
determining the baseline activity and the width of head direction
tuning.

The firing rate of each unit $i$ is determined through a
threshold-nonlinear transfer function,
\begin{equation*}
  \Psi_i^t = \Psi_{\rm sat} \, \tan^{-1}[g^t (\alpha_i^t - \mu^t)]
  \, \Theta(\alpha_i^t - \mu^t),
\end{equation*}

\noindent where $\Psi_{\rm sat} = 2/\pi$ normalizes the firing
rate into arbitrary units, $\Theta(\cdot)$ is the Heaviside
function, and $g^t$ and $\mu^t$ are the gain and threshold of the
nonlinearity, respectively. The variable $\alpha_i^t$ represents a
forgetful integration of the input $h_i$,
\begin{equation*}
  \alpha_i^t = \alpha_i^{t-1} + b_1(h_i^{t-1}-\beta_i^{t-1}-\alpha_i^{t-1}),
\end{equation*}

\noindent adapted by the input-dependent dynamical threshold
\begin{equation*}
  \beta_i^t = \beta_i^{t-1} + b_2(h_i^{t-1}-\beta_i^{t-1}),
\end{equation*}

\noindent where $\beta_i$ has a slower dynamics than $\alpha_i$,
$b_2 = b_1/3$ with $b_1 = 0.200$. Across the population of $N$
grid units, the mean activity $a = \sum_i \Psi_i^t / N$ and the
sparsity $s = (\sum_i \Psi_i^t)^2 / [N \sum_i (\Psi_i^t)^2]$ are
kept within $10\%$ relative to pre-specified values, $a_0 = 0.1$
and $s_0 = 0.3$, respectively, by appropriate temporal update of
the parameters $g^t$ and $\mu^t$ \cite{SiTreves2013}.

The feedforward and collateral connections play a key role in the
development of the grid scale and orientation alignment,
respectively, although crossed influences are also relatively
important. Feedforward connections $W_{ij}$ are learnt from random
initialization by Hebbian association,
\begin{equation*}
  W_{ij}^t = \left[W_{ij}^{t-1} + \epsilon\,(\Psi_i^t \, r_j^t -
  \bar{\Psi}_i^{t-1} \, \bar{r}_j^{t-1})\right]^+,
\end{equation*}

\noindent where $[\cdot]^+$ is the threshold function ($[x]^+ = 0$
for $x < 0$, and $[x]^+ = x$ otherwise), $\bar{\Psi}_i^{t-1}$ and
$\bar{r}_j^{t-1}$ are the time-dependent mean activities from grid
unit $i$ and place cell $j$, respectively, and $\epsilon = 0.005$
is a moderately low learning rate. After updating, these weights
are normalized according to $\sum_j (W_{ij}^t)^2 = 1$, which
mimics a homeostatic control of the synaptic function. Collateral
connections $\tilde{W}_{ik}$ are implemented {\em ad hoc},
putatively as the result of a long learning process taking place
in conjunctive layers of the mEC \cite{SiTreves2013}. Here, the
structure of these connections is formulated simply as the
extension of previous studies to the present topology
\cite{KropffTreves2008, Si_etal2012}. Each unit $i$ embedded in a
network with collaterals, having head-direction properties defined
by $f_{\theta_i}(\vartheta)$, is nominally associated to an
auxiliary field with a randomly chosen preferred location in the
bi-dimensional half-plane $(u_i,v_i)$. The collateral weight from
unit $k$ to unit $i$ is calculated as
\begin{equation*}
  \tilde{W}_{ik} = \left[ f_{\theta_k}(\vartheta_{ki}^{k}) \,
  f_{\theta_i}(\vartheta_{ki}^{i})\, \exp[-d_{ki}^2 / (2\sigma_f^2)]
  - \kappa \right]^+,
\end{equation*}

\noindent where $[\cdot]^+$ is the threshold function introduced
above, $\kappa$ is an inhibition parameter controlling sparseness
of the connections, $\vartheta_{ki}^{k}$ and $\vartheta_{ki}^{i}$
are the angles of the geodesics that join $(u_k,v_k)$ and
$(u_i,v_i)$, measured at the first and the second point,
respectively (note that, contrary to the Euclidean case, angles of
a given geodesics depend on the measurement point), $\sigma_f =
10$ cm is the spatial tuning, and
\begin{equation*}
  d_{ki} = \sqrt{\left[u_i-(u_k+\Delta u)\right]^2 + \left[v_i-(v_k+\Delta v)\right]^2},
\end{equation*}

\noindent where $\Delta u$ ($\Delta v$) is the distance on the
$u$-axis ($v$-axis) transversed from $(u_k,v_k)$ along the
geodesics to $(u_i,v_i)$ by a movement of $l = 10$ cm,
representative of the displacement carried out by the rat during
the delay period. As before, normalization of the weights is set
by $\sum_k (\tilde{W}_{ik})^2 = 1$.

\subsection{Characterization of local grid structure}
The triangular tile is the minimal structure associated to regular
hyperbolic tessellations. The two properties defining any regular
triangle are the length of the side and the internal angle.
Therefore, to characterize the local structure of the grid pattern
in an individual unit we extract these two properties from the
spikes it produces. First, we collect a representative number of
spike pairs in the half plane, eg $10^5$, to construct the
distribution of distances, ie along the geodesics joining both
locations. Typically, this distribution is highly multi-peaked,
where the first peak corresponds to distances between intra-field
spikes, the second peak between spikes belonging to neighbouring
fields, and subsequent peaks between spikes in non-adjacent
fields. Since the length of the side of the tiling triangle in a
regular pattern would correspond to the location of the second
peak, we define a range of distances around this peak as a filter
condition to declare spikes belonging to neighbouring fields. The
limits of this range were defined by the surrounding troughs, if
they exist, or fixed to $0.5\,d$ and $1.4\,d$, if they don't,
where $d$ is the distance corresponding to the second peak,
declared as the grid distance of the unit. For pseudo-spikes
randomly associated to visited locations, distances between spikes
are unimodally distributed, and hereafter utilized as a control
condition. Secondly, triplets of spikes were putatively classified
as belonging to neighbouring fields based on distance filtering in
the previous range, and the three internal angles determined
(under the underlying topology). These three angles were pooled
together and accumulated in an overall angular distribution
(obtained from $10^5$ effective triplets). The distribution of
angles so obtained for the spiking activity and the control
condition were different and their ratio was used to characterize
the angle subtended in the triangular pattern. Typically (in the
asymptotical state), this ratio was unimodal and distributed
asymmetrically around a peak. We defined the characteristic angle
as the median of the above-chance distribution (ratio values above
unity indicate an above-chance condition or, in other words,
angles more frequently obtained than chance).

\bigskip
\bigskip

\noindent \large{\bf Acknowledgments}

\bigskip

\normalsize This work was supported by the EU FET project GRIDMAP
(FP7-ICT 600725). Extensive discussions with other participants in
the consortium are gratefully acknowledged.


\begin{thebibliography}{10}

\bibitem{Khayyam1077}
Khayyam O. 1077 Explanations of the difficulties in the postulates
in Euclid's Elements. In english in: Amir-M\'oez A. 1959
Discussion of difficulties in Euclid. \textit{Scripta Mathematica}
{\bf 24}, 275-303.

\bibitem{Rozenfeld1988}
Rozenfeld BA. 1988 A history of non-Euclidean geometry: Evolution
of the concept of a geometric space. Berlin: Springer-Verlag.

\bibitem{Bonola1912}
Bonola R. 1912 Non-Euclidean geometry: A critical and historical
study of its development. Chicago: The Open Court Publishing
Company.

\bibitem{Halsted1900}
Halsted GB. 1900 Gauss and the non-Euclidean geometry. \textit{The
American Mathematical Monthly} {\bf 7}, 247-252.

\bibitem{Fyhn_etal2004}
Fyhn M, Molden S, Witter MP, Moser EI, Moser MB. 2004 Spatial
representation in the entorhinal cortex. \textit{Science} {\bf
305}, 1258-1264. (doi:10.1126/science.1099901)

\bibitem{Hafting_etal2005}
Hafting T, Fyhn M, Molden S, Moser MB, Moser EI. 2005
Microstructure of a spatial map in the entorhinal cortex.
\textit{Nature} {\bf 436}, 801-806. (doi:10.1038/nature03721)

\bibitem{Moser_etal2008}
Moser EI, Kropff E, Moser MB. 2008 Place cells, grid cells, and
the brain's  spatial representation system. \textit{Annu Rev
Neurosci} {\bf 31}, 69-89.
(doi:10.1146/annurev.neuro.31.061307.090723)

\bibitem{Solstad2009}
Solstad T (2009) Neural representations of Euclidean space. PhD
Thesis (Norwegian University of Science and Technology,
Trondheim).

\bibitem{Fyhn_etal2007}
Fyhn M, Hafting T, Treves A, Moser MB, Moser EI. 2007 Hippocampal
remapping and grid realignment in entorhinal cortex.
\textit{Nature} {\bf 446}, 190-194. (doi:10.1038/nature05601)

\bibitem{MoserMoser2008}
Moser EI, Moser MB. 2008 A metric for space. \textit{Hippocampus}
{\bf 18}, 1142-1156. (doi:10.1002/hipo.20483)

\bibitem{Barry_etal2007}
Barry C, Hayman R, Burgess N, Jeffery KJ. 2007
Experience-dependent rescaling of entorhinal grids. \textit{Nat
Neurosci} {\bf 10}, 682-684. (doi:10.1038/nn1905)

\bibitem{Stensola_etal2012}
Stensola H, Stensola T, Solstad T, Froland K, Moser MB, Moser EI.
2012 The entorhinal grid map is discretized. \textit{Nature} {\bf
492}, 72-78. (doi:10.1038/nature11649)

\bibitem{Fyhn_etal2008}
Fyhn M, Hafting T, Witter MP, Moser EI, Moser MB. 2008 Grid cells
in mice. \textit{Hippocampus} {\bf 18}, 1230-1238.
(doi:10.1002/hipo.20472)

\bibitem{Yartsev_etal2011}
Yartsev MM, Witter MP, Ulanovsky N. 2011 Grid cells without theta
oscillations in the entorhinal cortex of bats. \textit{Nature}
{\bf 479}, 103-107. (doi:10.1038/nature10583)

\bibitem{Hayman_etal2011}
Hayman R, Verriotis MA, Jovalekic A, Fenton AA, Jeffery KJ. 2011
Anisotropic encoding of three-dimensional space by place cells and
grid cells. \textit{Nat Neurosci} {\bf 14}, 1182-1188.
(doi:10.1038/nn.2892)

\bibitem{Jeffery_etal2013}
Jeffery KJ, Jovalekic A, Verriotis M, Hayman R. 2013 Navigating in
a three-dimensional world. \textit{Behav Brain Sci} {\bf 36},
523-587. (doi:10.1017/S0140525X12002476)

\bibitem{YartsevUlanovsky2013}
Yartsev MM, Ulanovsky N. 2013 Representation of three-dimensional
space in the hippocampus of flying bats. \textit{Science} {\bf
340}, 367-372. (doi:10.1126/science.1235338)

\bibitem{Ulanovsky_Fens2014}
Ginosar G, Finkelstein A, Fellous JM, Las L, Ulanovsky N. 2014 In
search of 3-D grid cells in flying bats. \textit{Fens 2014 Book of
Abstracts}, FENS-2107.

\bibitem{Stella_etal2013a}
Stella F, Si B, Kropff E, Treves A. 2013 Grid maps for
spaceflight, anyone? They are for free! \textit{Behav Brain Sci}
{\bf 36}, 566-567. (doi:10.1017/S0140525X13000575)

\bibitem{Beltrami1868}
Beltrami E. 1868 Saggio di interpretazione della geometria non
euclidea. \textit{Gior Mat} (in italian) {\bf 6}, 248-312.

\bibitem{Stella_etal2013b}
Stella F, Si B, Kropff E, Treves A. 2013 Grid cells on the ball.
\textit{J Stat Mech} {\bf P03013}.
(doi:10.1088/1742-5468/2013/03/P03013)

\bibitem{KropffTreves2008}
Kropff E, Treves A. 2008 The emergence of grid cells: Intelligent
design or just adaptation? \textit{Hippocampus} {\bf 18},
1256-1269. (doi:10.1002/hipo.20520)

\bibitem{Kruge_etal2013}
Kruge IU, Wernle T, Moser EI, Moser MB. 2013 Grid cells of animals
raised in spherical environments. \textit{Soc Neurosci Abstr} {\bf
39}, 769.14.

\bibitem{Saccheri1733}
Saccheri H. 1733 Euclides ab omni naevo vindicatus. Available in
http://mathematica.sns.it/opere/128/.

\bibitem{Si_etal2012}
Si B, Kropff E, Treves A. 2012 Grid alignment in entorhinal
cortex. \textit{Biol Cybern} {\bf 106}, 483-506.
(doi:10.1007/s00422-012-0513-7)

\bibitem{Sargolini_etal2006}
Sargolini F, Fyhn M, Hafting T, McNaughton BL, Witter MP, Moser
MB, Moser EI. 2006 Conjunctive representation of position,
direction, and velocity in entorhinal cortex. \textit{Science}
{\bf 312}, 758-762. (doi:10.1126/science.1125572)

\bibitem{SiTreves2013}
Si B, Treves A. 2013 A model for the differentiation between grid
and conjunctive units in medial entorhinal cortex.
\textit{Hippocampus} {\bf 23}, 1410-1424. (doi:10.1002/hipo.22194)

\bibitem{Burgess_etal2007}
Burgess N, Barry C, O'Keefe J. 2007 An oscillatory interference
model of grid cell firing. \textit{Hippocampus} {\bf 17}, 801-812.
(doi:10.1002/hipo.20327)

\bibitem{Blair_etal2007}
Blair HT, Welday AC, Zhang K. 2007 Scale-invariant memory
representations emerge from Moir\'e interference between grid
fields that produce theta oscillations: A computational model.
\textit{J Neurosci} {\bf 27}, 3211-3229.
(doi:10.1523/JNEUROSCI.4724-06.2007)

\bibitem{McNaughton_etal2006}
McNaughton BL, Battaglia FP, Jensen O, Moser EI, Moser MB. 2006
Path integration and the neural basis of the `cognitive map'.
\textit{Nat Rev Neurosci} {\bf 7}, 663-678. (doi:10.1038/nrn1932)

\bibitem{Dillon_etal2013}
Dillon MR, Huang Y, Spelke ES. 2013 Core foundations of abstract
geometry. \textit{Proc Natl Acad Sci USA} {\bf 110}, 14191-14195.
(doi:10.1073/pnas.1312640110)

\bibitem{Dehaene_etal2006}
Dehaene S, Izard V, Pica P, Spelke E. 2006 Core knowledge of
geometry in an Amazonian indigene group. \textit{Science} {\bf
311}, 381-384. (doi:10.1126/science.1121739)

\bibitem{Izard_etal2011}
Izard V, Pica P, Spelke ES, Dehaene S. 2011 Flexible intuitions of
Euclidean geometry in an Amazonian indigene group. \textit{Proc
Natl Acad Sci USA} {\bf 108}, 9782-9787.
(doi:10.1073/pnas.1016686108)

\bibitem{Tonnevie_etal2013}
Bonnevie T, Dunn B, Fyhn M, Hafting T, Derdikman D, Kubie JL,
Roudi Y, Moser EI, Moser MB. 2013 Grid cells require excitatory
drive from the hippocampus. \textit{Nat Neurosci} {\bf 16},
309-317. (doi:10.1038/nn.3311)

\bibitem{Rowland_etal2013}
Rowland DC, Weible AP, Wickersham IR, Wu H, Mayford M, Witter MP,
Kentros CG. 2013 Transgenically targeted rabies virus demonstrates
a major monosynaptic projection from hippocampal area CA2 to
medial entorhinal layer II neurons. \textit{J Neurosci} {\bf 33},
14889-14898. (doi:10.1523/JNEUROSCI.1046-13.2013)

\bibitem{Zhang1996}
Zhang K. 1996 Representation of spatial orientation by the
intrinsic dynamics of the head-direction cell ensemble: A theory.
\textit{J Neurosci} {\bf 16}, 2112-2126.

\end{thebibliography}
\end{document}